\documentclass[11pt]{article}
\newcommand{\X}{{\scriptscriptstyle X}}
\date{\small August 24$^\mathrm{th}$ 1998}
\title{Constraint on the Isgur-Wise function for $\overline
  B\rightarrow D$ semileptonic decays} 
\author{Antonio O.\ Bouzas \thanks{E-mail:
    abouzas@kin.cieamer.conacyt.mx} \hspace{5ex} Virendra Gupta 
  \\\small Departamento de F\'{\i}sica
  Aplicada, CINVESTAV-IPN \\\small Apdo.\ Postal 73 ``Cordemex,''
  M\'erida 97310, Yucat\'an, M\'exico}

\begin{document}

\maketitle

\begin{abstract}
  A current algebra sum rule for the inclusive decays $\overline
  B\longrightarrow D+X+\ell+\overline\nu$ by Bjorken \emph{et al.} is
  used to obtain constraints on the Isgur-Wise function
  $\xi(\omega)$ for the $\overline B\longrightarrow D
  \ell\overline\nu$ decays.
\end{abstract}

In Heavy Quark Effective Theory (HQET) the calculation of
weak current 
matrix elements of hadrons  which contain heavy quarks is greatly
simplified \cite{vo}--\cite{fa}. In general, six form factors determine the
hadronic matrix elements in the semileptonic decays
\begin{equation}
  \label{decay}
  \overline B(p)\longrightarrow D(p^\prime)\ell\overline\nu 
  \hspace{3ex}\mbox{and}\hspace{3ex} 
  \overline B(p)\longrightarrow D^*(p^\prime)\ell\overline\nu. 
\end{equation}
In HQET, all form factors for these decays are expressed
in terms of one unknown scalar function $\xi(\omega)$, the Isgur-Wise
function.  The argument of $\xi$ is the Lorentz scalar
$\omega=v\cdot v^\prime$, where $v_\mu=p_\mu/m_B$,
$v^{\prime\mu}=p^\prime_\mu/m_D$  are the four velocities of $\overline B$ and
$D$ (or $D^*$) mesons.

Using the equal time commutators of the heavy quark currents,
Bjorken \emph{et al.}\cite{bj} have derived sum rules for bottom baryons
and mesons.  We use one of their sum rules for inclusive
decays   $\overline B\rightarrow (D \mbox{ or } D^*) X
\ell\overline\nu$ to obtain constraints on the Isgur-wise function
$\xi(\omega)$ for semileptonic decays.

For the $b$ and $c$ heavy-quark currents
$J^{(+)}=\overline c\Gamma b$ 
and $J^{(-)}=\overline b\,\overline\Gamma c$ the equal time commutator
is
\begin{eqnarray}
  [J^{(-)}(x),J^{(+)}(y)] & = & \left(\overline
  b\,\overline\Gamma\gamma_0\Gamma b - \overline
  c\Gamma\gamma_0\overline \Gamma c\right)
  \delta^3(\mathbf{x}-\mathbf{y}) \nonumber\\
  \label{etc}
  & \equiv & J^{(3)}(x) \delta^3(\mathbf{x}-\mathbf{y})
\end{eqnarray}
where $\Gamma$ is any Dirac matrix and $\overline\Gamma = \gamma_0
\Gamma^\dagger \gamma_0$. For arbitrary $\mathbf{q}$ Eq.\ (\ref{etc}) gives,
\begin{equation}
  \label{etc2}
  \left[\int\!d^3\!x\; e^{i\mathbf{q\cdot x}} J^{(-)}(x), 
    \int\!d^3\!x\; e^{-i\mathbf{q\cdot x}} J^{(+)}(y)\right] = 
  \int\!d^3\!x\; J^{(3)}(x).
\end{equation}
Taking matrix elements between $|\overline B(p)\rangle$ and
$|\overline B(p^\prime)\rangle$ states, and inserting a complete set
of intermediate states one obtains,
\begin{eqnarray}
 (2\pi)^3 \sum_n \left\{\left|\langle n | J^{(+)}(0) | \overline
  B(p)\rangle \right|^2 \delta^3(\mathbf{p-p_n-q})\right.  & & \nonumber\\ 
\left.-\left|\langle n | J^{(-)}(0) | \overline
  B(p)\rangle \right|^2 \delta^3(\mathbf{p-p_n+q})\right\} 
& = & 
 \langle \overline B(p) | J^{(3)}(0) | \overline B(p)\rangle.
  \label{sr}
\end{eqnarray}
The states $|n\rangle$ in the first term are of the form $|(D
\mbox{ or } D^*) X\rangle$, and in the second $|\overline B\,\overline
B (\overline D \mbox{ or } \overline D^*) X\rangle$, where $X$ can be
$\pi,$ $2\pi,$ etc. These two contributions correspond to all
old-fashioned perturbation theory diagrams for the weak-current
squared matrix elements, with the exception of ``bubble'' graphs
\cite{bj}.  In the heavy-quark limit the form
of the matrix elements in Eq.\ (\ref{sr}) is fixed by heavy-quark spin
symmetry, Lorentz covariance and parity conservation \cite{fa,bj}, up
to an unknown matrix function $\rho_\X(v,v^\prime)$.  A
detailed kinematical analysis \cite{bj} then shows that each term on
the left-hand side of Eq.\ (\ref{sr}) must satisfy a sum rule.  The
first one, in particular, leads to  \cite{bj} ($\omega\equiv
v\cdot v^\prime$)\footnote{We generally follow the notation
  used in \cite{bj} except for their $\omega(\varepsilon, v\cdot
  v^\prime)$, which we denote by $\Omega(\varepsilon, v\cdot
  v^\prime)$.}
\begin{equation}
  \label{sr2}
  \xi^2 (\omega) \frac{1+\omega}{2} + \int^\infty_0\!\! d\varepsilon\,
  \Omega(\varepsilon, \omega) = 1.  
\end{equation}
The first term in Eq.\ (\ref{sr2}) corresponds to the semileptonic decay
in Eq.\ (\ref{decay}), whose hadronic 
matrix elements of the weak currents $V_\mu$ or $A_\mu$ are all given in
terms of $\xi(\omega)$.\footnote{This term can also be obtained \cite{gu} by
keeping only the allowed one-particle intermediate states ($D$
and $D^*$) in Eq.\ (\ref{sr}), using standard current algebra techniques
and taking the heavy-quark limit afterwards.} The density
$\Omega$ takes the form \cite{bj}, 
\begin{equation}
  \label{omega}
  \Omega(\varepsilon,\omega) = \frac{1}{2} \sum_{X} \mathrm{Tr}\left(
  \Lambda_+(v) \rho_\X\Lambda_+(v^\prime)\overline\rho_\X \right)
  \delta(\varepsilon-v^\prime\cdot p_\X)
\end{equation}
where $\Lambda_+(v)=(1+\not\! v)/2$ and $\rho_\X(v,v^\prime)$ contains
the dynamics of the light degrees of freedom $|X\rangle,$ with four-momentum
$p_\X$ ($\overline\rho_\X=\gamma_0\rho_\X^\dagger\gamma_0).$  
The integration variable $\varepsilon$ in Eq.\ (\ref{sr2}) is defined
by $(m_D + \varepsilon)^2 = (p^\prime + p_\X)^2$, and in the heavy
quark limit it takes the form $\varepsilon = v^\prime\cdot p_\X +
O(m_D^{-1}),$  as explicitly shown in Eq.\ (\ref{omega}).

Since all terms in the expression (\ref{omega}) for $\Omega$ are
positive definite, from Eq.\ (\ref{sr2}) we can infer that,
\begin{equation}
  \label{sr3}
  \xi^2 (\omega) \frac{1+\omega}{2} + \int^{\cal E}_0\!\! d\varepsilon\,
  \Omega_\pi(\varepsilon, \omega) \leq 1.  
\end{equation}
Here, we have retained only the term $\Omega_\pi$ in Eq.\ 
(\ref{omega}) corresponding to $X=\pi,$ with
$0\leq\varepsilon\leq{\cal E}$ for some upper limit ${\cal E}$.
Dropping the remaining positive contributions to sum rule (\ref{sr2})
gives rise to the inequality sign in Eq.\ (\ref{sr3}). This inequality
is more stringent on $\xi(\omega)$ the larger we take ${\cal E}$ to be,
whereas for ${\cal E}=0$ it reduces to the well-known kinematical
upper-bound on $\xi(\omega)$ \cite{bj}.

For ${\cal E}$ of the order of the hadronic scale, 
${\cal E} \raisebox{-0.8ex}{$\stackrel{\displaystyle <}{\sim}$} 1$ GeV,
we can compute $\Omega_\pi$ in Heavy Hadron 
Chiral Perturbation Theory \cite{wi,bu,ya}.  The decay amplitudes at
tree level are given by,
\begin{eqnarray}
  \label{amp}
  \langle D_{f^\prime} (v^\prime) \pi^a(p_\pi)|J(0)|\overline
  B_f(v)\rangle & = & 
  \mathrm{Tr}\left[
  \Lambda_+(v^\prime)\Gamma\Lambda_+(v)\rho_\pi\right]\\ 
  \label{amp*}
  \langle D^*_{f^\prime} (v^\prime,\epsilon)
  \pi^a(p_\pi)|J(0)|\overline
  B_f(v)\rangle & = & 
  \mathrm{Tr}\left[\gamma_5\not\!\epsilon
  \Lambda_+(v^\prime)\Gamma\Lambda_+(v)\rho_\pi
  \right]\\
  \label{rhopi}
  \rho_\pi (v,v^\prime) & = & \frac{i g}{f_0}
  \tau^a_{f f^\prime}
  \xi(\omega)\left(\frac{v^\prime\cdot p_\pi + \not\! 
  p_\pi}{v^\prime\cdot p_\pi} + \frac{v\cdot p_\pi - \not\!
  p_\pi}{v\cdot p_\pi}\right)\gamma_5.
\end{eqnarray}
Note that $\rho_\pi$ is the same for the two amplitudes due to 
heavy-quark spin symmetry, and that $\xi (\omega)$ is the same
function as in the first term of Eqs.\ (\ref{sr2}) and (\ref{sr3}) due
to chiral 
symmetry.  In Eqs.\ (\ref{amp})--(\ref{rhopi}), $f$, $f^\prime$ and
$a$ are 
isospin indices for $\overline B$, $D\mbox{ or } D^*$, and $\pi$,
respectively, and $\tau^a$ are generators of the $su(2)$ isospin
algebra normalized to $\mathrm{Tr}(\tau^a\tau^b)=\delta^{ab}$.  $g$
is the 
coupling constant for the strong vertices  
$D^* D\pi$  and $B^* B\pi$,  and $f_0=\sqrt{2} f_\pi$ where $f_\pi =
93$ MeV is the pion decay constant.   
We have neglected $m_\pi$ and the mass difference between $D^*$
and $D$ mesons.  If we set $\Gamma = \gamma^\mu (1-\gamma_5),$ 
equations (\ref{amp}) and (\ref{amp*}) give a compact expression for
the 
weak decay amplitudes derived in \cite{ya}.

For $\Omega_\pi$ we then obtain the expression,\footnote{%
  $\Omega_\pi$ vanishes at the zero-recoil point $\omega =
  1$ due to the factor in square brackets in (\ref{omegalog}), a
  result which holds true also in the case of many pion emission
  \cite{ya2}. }
\begin{eqnarray}
  \Omega_\pi (\varepsilon, \omega) & = & \frac{1}{2} 
  \int\! \frac{d^4p_\pi}{(2\pi)^3} \delta(p_\pi^2)
  \delta(\varepsilon-v^\prime\cdot p_\pi)
  \mbox{Tr}(\Lambda_+(v)\rho_\pi
  \Lambda_+(v^\prime)\overline\rho_\pi)\nonumber\\
  & = & \frac{3}{8\pi^2} \frac{g^2}{f_0^2} 
  \varepsilon \xi^2(\omega)
  \left[\omega-\frac{1}{\sqrt{\omega^2-1}}
  \log\left(\omega+\sqrt{\omega^2-1}\right)\right]. 
  \label{omegalog}
\end{eqnarray}
Thus, inequality (\ref{sr3}) can be explicitly written as,
\begin{eqnarray}
  \label{xi2}
  \xi(\omega) & \leq & F(\omega)\equiv \left[\frac{1+\omega}{2} + \alpha
    \left(\omega-\frac{1}{\sqrt{\omega^2-1}}
    \log\left(\omega+\sqrt{\omega^2-1}\right)
\right)\right]^{-\frac{1}{2}}\\
  \label{alpha}
  \alpha & = & \frac{3}{16\pi^2} \frac{g^2{\cal E}^2}{f_0^2}~~.
\end{eqnarray}
The value of ${\cal E}$ must meet the requirement that contributions
to $F(\omega)$ from higher order corrections in the perturbative
chiral expansion must be small.  If, as we expect, the expansion
parameter is $g{\cal E}/(4\pi f_0),$ a value of ${\cal E} =$ 0.5 GeV
should be appropriate.  In fact, given the relative smallness of $g,$
such choice may be somewhat conservative.

Setting ${\cal E} = 0.5$ GeV yields $\alpha = 0.275 g^2.$  There is a
recent determination of $g$ from $D^*\rightarrow D\pi$ decay data \cite{st},
\renewcommand{\arraystretch}{0.5}
$g=0.27\begin{array}{l}\scriptstyle +0.04+0.05\\\scriptstyle -0.02-0.02
\end{array}$,%
\renewcommand{\arraystretch}{1}%
which leads to $\alpha = 0.020$.  (Note, however, that larger values
of $g$ are not completely ruled out by current data \cite{st}.) The
function $F(\omega)$ is plotted in Fig.\ 1 for several values of
$\alpha$. 

The derivative of $F(\omega)$ at zero recoil is given by
\begin{equation}
  \label{F}
  F^\prime(\omega=1) =
  -\frac{1}{4}\left(1+\frac{8\alpha}{3}\right).   
\end{equation}
Since $\xi^\prime(\omega=1)\leq F^\prime(\omega=1)$, for $\alpha=0$ we
recover the well-known upper bound $\xi^\prime (1)\leq -1/4$
\cite{bj}.  Inclusion of the low-lying $D\pi$ and $D^*\pi$ states
gives an improved upper bound which depends on $\alpha$.  For
$\alpha=0.020$, we get $\xi^\prime (1)\leq -0.263$.  To make this
inequality tighter one needs to include other multiparticle
states.

\subsection*{Acknowledgements}

One of us (V.G.) is thankful to Prof.\ Mark Wise for a discussion. We
thank the first referee for pointing out ref.\ \cite{ya2} to us.

\subsection*{Figure Caption}

The function $F(\omega)$ for $\alpha =$ 0 (solid line), 0.02 (short
dashes) and 0.08 (long dashes).  These values of $\alpha$ result from
setting $g=0.27$ (see main text) and ${\cal E} =$ 0, 0.5 and 1 GeV,
respectively. 

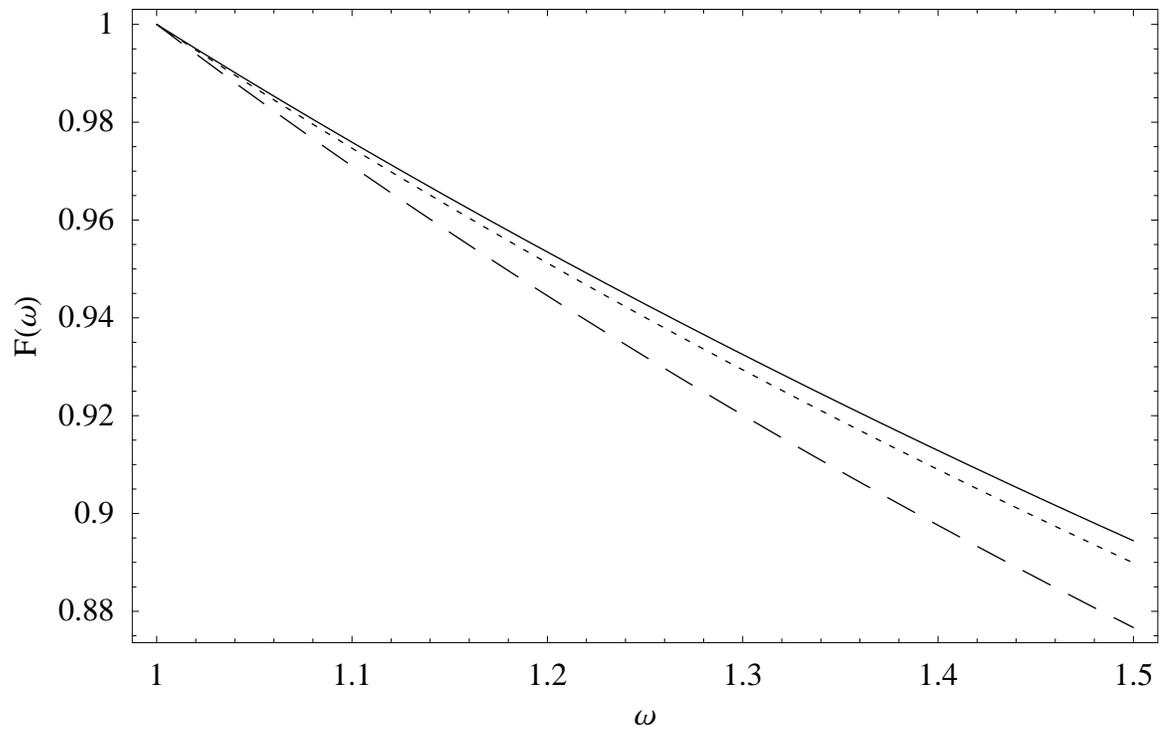
\end{document}